\documentclass [12pt]{article}
\usepackage [cp1251]{inputenc}
\usepackage [english]{babel}
\usepackage {graphicx}
\usepackage {amssymb}
\usepackage {amsmath}
\usepackage {longtable}
\title{Chaotic magnetoconvection in a non-uniformly rotating  electroconductive fluids}
 
\author{$^{1}$\textbf{M.I. Kopp}, $^3$\textbf{A.V. Tur}, $^{1,2}$\textbf{V.V. Yanovsky}}

\begin{document}

\maketitle

$^{1}$ \textit{Institute for Single Crystals, NAS  Ukraine, Nauky Ave. 60, Kharkov 61001, Ukraine}

$^{2}$\textit{V.N. Karazin Kharkiv National University 4 Svobody Sq., Kharkov 61022, Ukraine}

$^{3}$\textit{Universit\'{e} de Toulouse [UPS], CNRS, Institut de Recherche en Astrophysique et Plan\'{e}tologie,
9 avenue du Colonel Roche, BP 44346, 31028 Toulouse Cedex 4, France}

\abstract{We study a new type of magnetoconvection in a nonuniform  rotating  plasma layer under a constant vertical magnetic field. To describe the weakly nonlinear stage of convection we apply Galerkin-truncated approximation and we obtain the system of equations of  Lorentz  type. A numerical analysis of  these equations shows  the presence of chaotic behavior of convective flows. Criteria for the appearance of chaotic motions are found  depending on the convection parameters (Rayleigh number $\textrm{Ra}$), magnetic field (Chandrasekhar number $\textrm{Q}$), rotation (Taylor number $\textrm{Ta}$) for the Keplerian angular velocity profile $(\textrm{Ro}=-3/4)$ of the medium. 


\section{Introduction}

  Convective flows caused by thermal processes in the gravitational field are important for explaining many phenomena occurring in the bowels of planets, stars and other cosmic objects. Convection is the source of the generation of both large-scale magnetic fields and large-scale vortex structures in the laminar \cite{1s} or turbulent dynamo model \cite{2s}. Rotation and magnetic fields have a large effect on the convective processes of electroconductive media. The theory of such processes (the Rayleigh-Benard problem) for the case of uniform rotation and a constant magnetic field is described in detail in books \cite{3s}-\cite{4s}. 

  However, most of the various cosmic objects consisting of dense gases or fluids (Jupiter, Saturn, the Sun, Galaxies, etc.) and  the electrically conducting medium inside the planets, rotate non-uniformly. In many hydrodynamic problems, the differential rotation of the medium is modeled by the Couette flow enclosed between two cylinders, rotating at different angular velocities. This model is convenient for the realization of laboratory experiments. The stability of the Couette flow for an ideally conducting medium in a magnetic field was first considered in \cite{5s}-\cite{6s}.  It is shown, that a weak axial magnetic field destabilizes the azimuthal differential rotation of the plasma when the condition $d\Omega^{2}/dR<0$  is satisfied. As a result, in the non-dissipative plasma arises the  magneto rotational instability (MRI) or the standard magneto rotational instability (SMRI).

Since this condition is also satisfied for Keplerian flows  $\Omega \sim R^{-3/2} $ , the MRI is the most probable source of turbulence in accretion disks. The discovery of the MRI engenders  numerous theoretical studies. At the beginning this dealt with the problem of accretion flows in the approximation of the non-dissipative plasma with radial thermal stratification \cite{7s}, considering the magnetization of the heat fluxes \cite{8s}. In \cite{9s}, the stability of  differential-rotating plasma in an axial magnetic field is examined  with both dissipative effects (viscosity and ohmic dissipation) and thermal radial stratification of the plasma  as well. MRI in a helical magnetic field, i.e. with a nontrivial topology $\vec{B}_{0} rot\vec{B}_{0} \ne 0$  was studied in \cite{10s} -\cite{11s}. The model of rotating cylinders is used in the theory of convective dynamo (the Busse model), developed in \cite{12s}.

In this paper we investigate the weakly nonlinear stage of a non-uniformly rotating magnetoconvection in which a chaotic regime arises, leading to random variations of the magnetic field. Over the past few years, the chaotic behavior of convection has been intensively studied in rotating fluid layers \cite{13s}, in conducting media with a homogeneous magnetic field \cite{14s}, and also in conducting media rotating with a magnetic field \cite{15s}. However, these studies did not consider the dynamics of the magnetic field itself, which corresponds to a nonconductive approximation. This is  of great importance for astrophysics and for technological applications such as crystal growth, chemical processes of solidification and centrifugal casting of metals as well.

\section{Main equations}

To describe nonlinear convective phenomena in a non-uniformly rotating layer of an electroconductive fluid, it is convenient  to introduce a rotating frame of reference with local Cartesian coordinates $(x,y,z)$ (see Fig. \ref{fg1}). This frame of reference rotates with an angular velocity $\vec{\Omega }=\Omega (R)\vec{e}_{z} $, where cylindrical coordinates locally correspond to the Cartesian coordinates: $x$ -- the radial direction, $y$ -- the azimuthal axis  and $z$ -- the axial direction parallel to the rotation axis. The constant magnetic field $\vec{B}_{0} =\textrm{const}$   is assumed to be parallel to  the rotation axis: $\vec{B}_{0} \parallel OZ$. Consequently, the non-uniform   rotation of the fluid layer can be locally represented as a rotation with a constant angular velocity $\vec{\Omega }_{0} $ and azimuthal shear \cite{16s}, whose velocity profile is locally linear: $\vec{U}_{0} =-q\Omega _{0}x\vec{e}_{y} $, where  $q=-d\ln \Omega /d\ln R=3/2$  is a dimensionless broad parameter determined from the angular velocity profile of rotation: $\Omega (R)=\Omega_{0} (R/R_{0} )^{-q} $.  It is not difficult to see, that the shearing sheet parameter $q$  is related to the Rossby number $\textrm{Ro}=\frac{R}{2\Omega } \frac{\partial \Omega }{\partial R} $  by the relation:  $q=-2\textrm{Ro}$. 

The equations of magnetohydrodynamics in the Boussinesq approximation for perturbed quantities take the following form:
\begin{equation} \label{eq1}   \frac{{\partial \vec u}}{{\partial t}} -q {\Omega}_0 x\frac{\partial \vec u}{\partial y}  + (\vec u\nabla )\vec U_0 +2\vec{\Omega}\times \vec{u}+ (\vec u\nabla )\vec u  =  - \frac{1}{{\rho _0 }}\nabla \widetilde p + \frac{1}{{4\pi \rho _0 }}\left((\vec B_0 \nabla )\vec b+(\vec b\nabla )\vec b  \right) + g\beta \theta \vec{e}  + \nu \nabla^2 \vec u  \end{equation} 
\begin{equation} \label{eq2} \frac{{\partial \vec b}}{{\partial t}} -q {\Omega}_0 x\frac{\partial \vec b}{\partial y} - (\vec B_0 \nabla )\vec u - (\vec b\nabla )\vec U_0 +(\vec u\nabla )\vec b-(\vec b\nabla )\vec u = \eta \nabla^2 \vec b \end{equation}  
 \begin{equation} \label{eq3} \frac{{\partial \theta }}{{\partial t}} -q {\Omega}_0 x\frac{\partial \theta}{\partial y}  + (\vec u\nabla )T_0+(\vec u\nabla )\theta  = \chi \nabla^2 \theta \end{equation} 
\begin{equation} \label{eq4}
\textrm{div}\vec{b}=0, \quad \textrm{div}\vec{u}=0  \end{equation} 
$\nabla T_0$ is the constant temperature gradient between layer $T_1$ and $T_2$. Let us consider the dynamics of axisymmetric perturbations, when all perturbed quantities in equations (\ref{eq1})-(\ref{eq4}) will depend only on two variables $(x,z)$  : 
\[\vec{u}=(u,v,w),\; \vec{b}=(\widetilde u, \widetilde v, \widetilde w),\; \widetilde p=\widetilde p(x,z),\; \theta=\theta (x,z)\]
Solenoidal equations for axisymmetric perturbations of velocity and magnetic field take the form:
\begin{equation} \label{eq5} \frac{\partial u}{\partial x}+\frac{\partial w}{\partial z}=0, \quad \frac{\partial \widetilde u}{\partial x}+\frac{\partial \widetilde w}{\partial z}=0
\end{equation}	

\begin{figure}
  \centering
\includegraphics[width=7 cm, height=7 cm]{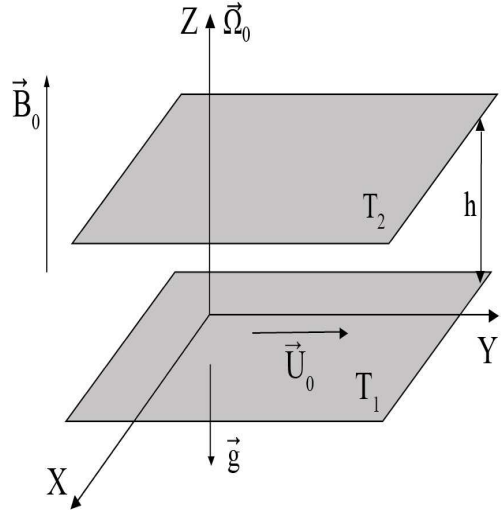}  \\
\caption{Physical configuration of the problem }\label{fg1}
\end{figure}

In accordance with the equations (\ref{eq5}), we can introduce the stream function  $\psi $ and the flux function $\phi $:
\[u=-\frac{\partial\psi}{\partial z},\quad w=\frac{\partial\psi}{\partial x},\quad \widetilde u=-\frac{\partial\phi}{\partial z},\quad \widetilde w=\frac{\partial\phi}{\partial x} .\]
We write down equations (\ref{eq1})- (\ref{eq3}) in terms of stream functions $\psi $ and $\phi $:

\begin{equation} \label{eq6} 
\frac{\partial }{{\partial t}}\nabla ^2 \psi  + 2\Omega _0 \frac{{\partial v}}{{\partial z}} - \frac{{B_0 }}{{4\pi \rho _0 }}\frac{\partial }{{\partial z}}\nabla ^2 \phi-g\beta \frac{{\partial \theta }}{{\partial x}} - \nu \nabla ^4 \psi  = \frac{1}{{4\pi \rho_0 }}J(\phi,\nabla^2 \phi )-J(\psi ,\nabla ^2 \psi ) \end{equation}
\begin{equation} \label{eq7} 
\frac{{\partial v}}{{\partial t}} - 2\Omega _0 (1 +\textrm{Ro})\frac{{\partial \psi }}{{\partial z}} - \frac{{B_0 }}{{4\pi \rho _0 }}\frac{{\partial \widetilde v}}{{\partial z}} - \nu \nabla ^2 v = \frac{1}{{4\pi \rho _0 }}J(\phi ,\tilde v) - J(\psi ,v)
 \end{equation}
\begin{equation} \label{eq8}
\frac{{\partial \phi }}{{\partial t}} - B_0 \frac{{\partial \psi }}{{\partial z}} - \eta \nabla ^2 \phi  =  - J(\psi ,\phi )\end{equation}
\begin{equation} \label{eq9}
\frac{{\partial \widetilde v}}{{\partial t}} - B_0 \frac{{\partial v}}{{\partial z}} + 2\Omega_0\textrm{Ro}\frac{{\partial \phi }}{{\partial z}} - \eta \nabla ^2 \widetilde v = J(\phi ,v)-J(\psi,\widetilde v)\end{equation}
\begin{equation} \label{eq10}
\frac{{\partial \theta }}{{\partial t}} - \frac{T_{1}-T_{2} }{h}\cdot\frac{{\partial \psi }}{{\partial x}} - \chi \nabla ^2 \theta  =  - J(\psi ,\theta )
\end{equation}
where $J(a,b)=\frac{\partial a}{\partial x} \frac{\partial b}{\partial z} -\frac{\partial a}{\partial z} \frac{\partial b}{\partial x} $ is the Jacobian or Poisson bracket $J(a,b)\equiv \left\{a,b\right\}$.
We note that in the absence of thermal phenomena, the system of equations (\ref{eq6}) - (\ref{eq10}) was obtained in article \cite{17s}. Since, here we consider the thermal phenomena, it is convenient in Eqs.(\ref{eq6}) - (\ref{eq10}) to go over to dimensionless variables:
\[ (x,z)=h(x^*,z^*), \; t=\frac{h^2}{\nu}t^*,\; \psi=\chi\psi^*,\; \phi=h B_0 \phi^*,\; v=\frac{\chi}{h}v^*,\; \widetilde v=B_0 \widetilde v^*,\; \theta=(T_{1}-T_{2} )\theta^* .\]
For simplicity let us omit asterisks. Then these equations in the dimensionless variables take the following form:
\begin{equation} \label{eq11}
\frac{\partial }{{\partial t}}\nabla ^2 \psi  + \sqrt {\textrm{Ta}} \frac{{\partial v}}{{\partial z}} - \Pr\textrm{Pm}^{-1}\textrm{ Q}\frac{\partial }{{\partial z}}\nabla ^2 \phi  -\textrm{Ra}\frac{{\partial \theta }}{{\partial x}} - \nabla ^4 \psi  = \Pr\textrm{Pm}^{-1}\textrm{Q}\cdot J(\phi ,\nabla ^2 \phi ) - {\Pr}^{-1}\cdot J(\psi ,\nabla ^2 \psi )\end{equation}
 \begin{equation} \label{eq12}
\frac{{\partial v}}{{\partial t}} - \sqrt {\textrm{Ta}} (1 +\textrm{Ro})\frac{{\partial \psi }}{{\partial z}} - \Pr\textrm{Pm}^{ - 1}\textrm{Q}\frac{{\partial \widetilde v}}{{\partial z}} - \nabla ^2 v = \Pr\textrm{Pm}^{ - 1}\textrm{Q}\cdot J(\phi ,\widetilde v) - {\Pr}^{-1}\cdot J(\psi ,v) \end{equation}

\begin{figure}
  \centering
\includegraphics[width=7 cm, height=7 cm]{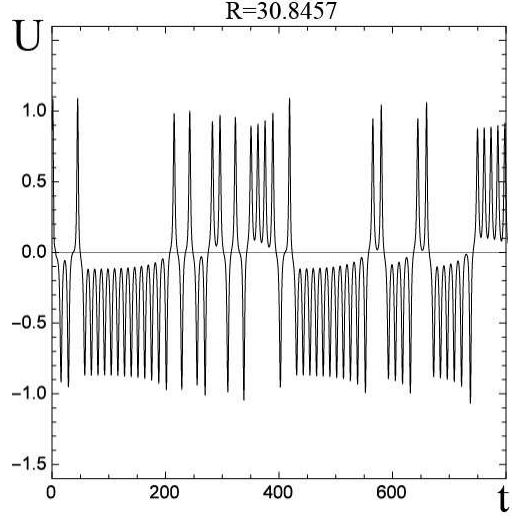}  \\
\caption{ Variations of the perturbed magnetic field. }\label{fg2}
\end{figure}

\begin{equation} \label{eq13}
 \frac{{\partial \phi }}{{\partial t}} - {\Pr} ^{-1} \frac{{\partial \psi }}{{\partial z}} -\textrm{Pm}^{-1} \nabla ^2 \phi  =  - {\Pr}^{-1} J(\psi ,\phi ))\end{equation}
\begin{equation} \label{eq14}
\frac{{\partial \widetilde v}}{{\partial t}} -{\Pr}^{-1} \frac{{\partial v}}{{\partial z}} +\textrm{Ro}\sqrt{\textrm{Ta}} \frac{{\partial \phi }}{{\partial z}} -\textrm{Pm}^{ - 1} \nabla ^2 \widetilde v = {\Pr}^{-1}(J(\phi ,v) - J(\psi ,\widetilde v)) \end{equation}
\begin{equation} \label{eq15}
\Pr \frac{{\partial \theta }}{{\partial t}} - \frac{{\partial \psi }}{{\partial x}} - \nabla ^2 \theta  =  - J(\psi ,\theta ))\end{equation}
where the dimensionless parameters are: $\textrm{ Pr}=\nu /\chi $  -- the Prandtl number, $\textrm{Pm}=\nu /\eta $ -- the magnetic Prandtl number, $\textrm{Ta}=\frac{4 \Omega_{0}^{2} h^{4} }{\nu ^{2} } $ -- the Taylor number,$ \textrm{Ha}=\frac{B_{0} h}{\sqrt{4\pi \rho_{0} \nu \eta } } $ -- the Hartmann number, $ \textrm{Ra}=\frac{g\beta (T_{1}-T_{2} )h^{3} }{\nu \chi } $ -- the Rayleigh number,  $\textrm{Q}=\textrm{Ha}^{2} $ -- the Chandrasekhar number.
The system of equations (\ref{eq11}) - (\ref{eq15}) is supplemented by the following boundary conditions:

\[\psi=\nabla^2\psi=0,\quad \frac{dv}{dz}=0,\quad \widetilde v=0, \quad \frac{d \phi}{dz}=0,\quad \theta=0 \quad \textrm{at} \quad z=0,\]
\begin{equation} \label{eq16} \end{equation}
\[\psi=\nabla^2\psi=0,\quad \frac{dv}{dz}=0,\quad \widetilde v=0, \quad \frac{d\phi}{dz}=0,\quad \theta=0 \quad \textrm{at} \quad z=1. \]

\section{Truncated Galerkin expansion}

To obtain the solution of nonlinear coupled system of partial differential Eqs.(\ref{eq11})-(\ref{eq15}), we use the Galerkin expansion in $x$ and $z$ -- directions for  perturbations:
\[ \psi(x,z,t)=A_1(t)\sin(kx)\sin(\pi z), \]
\[v=V_1(t)\sin(kx)\cos(\pi z),\]
\begin{equation}\label{eq17} \phi(x,z,t)=B_1(t)\sin(kx)\cos(\pi z),  \end{equation}
\[\widetilde{v}=W_1(t)\sin(kx)\sin(\pi z),\]
\[\theta(x,y,t)=C_1(t)\cos(kx)\sin(\pi z)+C_2(t)\sin(2\pi z), \]
where $k=2\pi h/L$ -- dimensionless wave number, $L$ -- the characteristic length of the layer in the horizontal direction, $A_{1} $, $V_{1} $, $B_{1} $, $W_{1} $, $C_{1} $, $C_{2} $ - perturbation amplitudes. 
\begin{figure}
  \centering
\includegraphics[width=7 cm, height=7 cm]{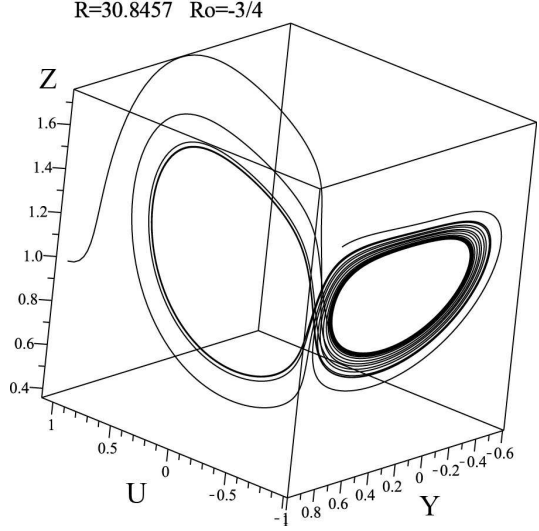}  \\
\caption{Three-dimensional projections of trajectories of chaotic motions. }\label{fg3}
\end{figure}
Substitute the expansion (\ref{eq17}) into the equations (\ref{eq11})-(\ref{eq15}). Then integrating them over the domain $[0,1]\times [0,L/h]$, taking into account orthogonality of functions we obtain the set of six ordinary differential equations for the time evolution of the amplitudes:
\[ \frac{\partial A_1}{\partial\widetilde{t}}=-A_1-\frac{\pi\sqrt{\textrm{Ta}}}{a^4}\cdot V_1- \frac{\pi\textrm{QPr}}{a^2\textrm{Pm}}\cdot B_1+\frac{k\textrm{Ra}}{a^4}\cdot C_1 \]
\[\frac{\partial V_1}{\partial\widetilde{t}}=-V_1+\frac{\pi\sqrt{\textrm{Ta}}}{a^2}(1+\textrm{Ro})\cdot A_1+\frac{\pi\textrm{QPr}}{a^2\textrm{Pm}}\cdot W_1 \]
\begin{equation}\label{eq18} \textrm{Pm}\frac{\partial B_1}{\partial\widetilde{t}}=-B_1+\frac{\pi\textrm{Pm}}{a^2\textrm{Pr}}\cdot A_1  \end{equation}
\[  \textrm{Pm}\frac{\partial W_1}{\partial\widetilde{t}}=-W_1-\frac{\pi\textrm{Pm}}{a^2\textrm{Pr}}\cdot V_1+\frac{\pi\textrm{PmRo}\sqrt{\textrm{Ta}}}{a^2}\cdot B_1   \]
\[  \textrm{Pr}\frac{\partial C_1}{\partial\widetilde{t}}=-C_1+\frac{k}{a^2}\cdot A_1+\frac{\pi k}{a^2}\cdot A_1C_2  \]
\[\textrm{Pr}\frac{\partial C_2}{\partial\widetilde{t}}=-\frac{4\pi^2}{a^2}\cdot C_2-\frac{\pi k}{2a^2}\cdot A_1C_1   \]
Here $a=\sqrt{k^{2} +\pi ^{2} } $, total wavelength number and time is rescaled by $\tilde{t}=a^{2} t$. So we obtain
the system of ordinary differential equations (\ref{eq18}) of  a low order spectral model, but it can fully reproduce convective processes in the complete  nonlinear system of equations (\ref{eq11}) - (\ref{eq15}). For convenience we introduce the following notation 
\[ \textrm{R}=\frac{k^2\textrm{Ra}}{a^6},\quad \textrm{T}=\frac{\pi^2\sqrt{\textrm{Ta}}}{a^6}, \quad \textrm{H}=\frac{\pi^2}{a^4}\frac{\textrm{QPr}}{\textrm{Pm}},\quad \gamma=\frac{4\pi^2}{a^2} \]
and we rescale the amplitudes  $A_{1} $, $V_{1} $, $B_{1} $, $W_{1} $, $C_{1} $, $C_{2} $  in the form:
\[X(\widetilde{t})=\frac{k\pi}{a^2\sqrt{2}}A_1(\widetilde{t}),\;V(\widetilde{t})=\frac{kV_1(\widetilde{t})}{\sqrt{2}},\; U(\widetilde{t})=\frac{kB_1(\widetilde{t})}{\sqrt{2}},\; W(\widetilde{t})=\frac{a^2k}{\pi\sqrt{2}}W_1(\widetilde{t}),\]
\[ Y(\widetilde{t})=\frac{\pi C_1(\widetilde{t})}{\sqrt{2}}, \; Z(\widetilde{t})=-\pi C_2(\widetilde{t}) \]

\begin{figure}
  \centering
\includegraphics[width=7 cm, height=7 cm]{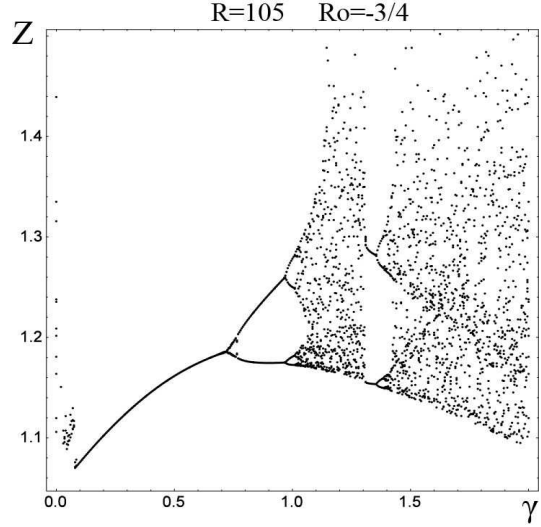}  \\
\caption{ Bifurcation diagram for  $Z$ -- amplitude  from parameter  changes. }\label{fg4}
\end{figure}

to obtain the following set of equations,
\begin{equation}\label{eq19}
\left\{
\begin{aligned}
\dot{X}=-X+\textrm{R}Y-\textrm{T}V-\textrm{H}U\\
\dot{V}=-V+\textrm{H}W+\sqrt{\textrm{Ta}}(1+\textrm{Ro})X\\
\dot{U}=-\textrm{Pm}^{-1}U+\textrm{Pr}^{-1}X \\
\dot{W}=-\textrm{Pm}^{-1}W-\textrm{Pr}^{-1}V+\textrm{Ro}\sqrt{\textrm{Ta}}U \\
\dot{Y}=\textrm{Pr}^{-1}(-Y+X-XZ) \\
\dot{Z}=\textrm{Pr}^{-1}(-\gamma Z+XY) \\
\end{aligned}
\right.
\end{equation}
where the dots $(.)$   denote the time derivative $\frac{d}{d\tilde{t}} $.  Eqs. (\ref{eq19}) are like the Lorenz equations \cite{18s}, but  only   for a six-dimensional phase space.

\section{ Stability analyses }

Qualitative and numerical analysis of Eqs. (\ref{eq19}) allows us to determine the type of fixed points and the conditions for  a chaotic regime. It is easy to see, that the system of equations (\ref{eq19}) is dissipative, since the divergence of the volume in phase space  is negative:
\[\textrm{div}\vec{\Phi}=\frac{\partial \dot{X} }{\partial X}+\frac{\partial \dot{V} }{\partial V}+\frac{\partial \dot{U} }{\partial U}+\frac{\partial \dot{W} }{\partial W}+\frac{\partial \dot{Y} }{\partial Y}+\frac{\dot{\partial Z} }{\partial Z}=-2(1+\textrm{Pm}^{-1})-\textrm{Pr}^{-1}(1+\gamma)<0 \]
Hence, if set of initial points in the phase space occupies the volume $\vec \Phi (0) $  at time $t=0$, then the volume in the phase space is
\[\vec{\Phi}(\widetilde{t})=\vec{\Phi}(0)\exp[\left(-2(1+\textrm{Pm}^{-1})-\textrm{Pr}^{-1}(1+\gamma)\right)\widetilde{t}]. \]
This expression shows that the volume decreases exponentially with time. Thus, in the phase space of dissipative systems appear  attractors. Moreover, the system of equations (\ref{eq19}) is invariant with respect to the substitution 
$(X,V,U,W,Y,Z)\to (-X,-V,-U,-W,-Y,Z)$. System of  Eqs. (\ref{eq19})  has the general form $\dot{X}_{s} =f(X_{s})$ and the equilibrium (fixed or stationary) points are obtained by $f(X_{s})=0$: 
\[(X_1,V_1,U_1,W_1,Y_1,Z_1)=(0,0,0,0),\]
\[\left(X_2,X_3\right)=\pm \frac{1}{r}\sqrt{\gamma r ( \textrm{R}-r)},\; \left(V_2,V_3\right)=\pm \frac{\sqrt{\textrm{Ta}}\left(\textrm{HRo}\textrm{Pm}^2+\textrm{Pr}(1+\textrm{Ro})\right)}{r(\textrm{HPm}+\textrm{Pr})}\sqrt{\gamma r (\textrm{R}-r)} ,  \]
\[\left(U_2,U_3\right)=\pm \frac{\textrm{Pm}}{r\textrm{Pr}} \sqrt{\gamma r (\textrm{R}-r)}, \; \left(W_2,W_3\right)=\pm \frac{\sqrt{\textrm{Ta}}\textrm{Pm}\left(\textrm{RoPm}-\textrm{Ro}-1\right)}{r(\textrm{HPm}+\textrm{Pr})} \sqrt{\gamma r (\textrm{R}-r)},  \]
\[ \left(Y_2,Y_3\right)=\pm \frac{1}{\textrm{R}}\sqrt{\gamma r (R-r)},\;\left(Z_2,Z_3\right)=1-\frac{r}{\textrm{R}}, \]
where
\[r= 1+\frac{\textrm{Pm}}{\textrm{Pr}}\textrm{H}+\textrm{T}\sqrt{\textrm{Ta}}\cdot\frac{1+\textrm{Ro}\left(1+\frac{\textrm{Pm}^2}{\textrm{Pr}}\textrm{H}\right)}{1+\frac{\textrm{Pm}}{\textrm{Pr}}\textrm{H}}.\]
To determine the type of fixed points, we linearize the system of equations (\ref{eq19}) in a small neighborhood of fixed points using the standard method.
As a result, we write the linearized equations in the form of a Jacobi matrix.  The characteristic values $\lambda_{i}$  $(i=1,2,3,4,5,6)$ of the Jacobian matrix, at the vanishing of  the characteristic polynomial 
\[P(\lambda)\equiv a_0\lambda^6+a_1\lambda^5+a_2\lambda^4+a_3\lambda^3+a_4\lambda^2+a_5\lambda+a_6=0,\quad a_0=1>0, \]
 provide the stability conditions. 

The explicit form of the real coefficients  $a_{1}, a_{2}, a_{3}, a_{4}, a_{5}, a_{6} $   is not given since their form is very cumbersome. However, we can use the Rauss-Hurwitz criterion known from the theory of asymptotic stability \cite{19s}. In order the polynomial  $P(\lambda )$  has all roots with negative real parts it is necessary and sufficient, that  the following conditions be satisfied: 
\begin{enumerate}
\item all the coefficients of the polynomial $P(\lambda )$  were positive  $a_{n} >0$, $n=1\div 6$  ;
\item the following inequalities are valid for the Hurwitz determinants: $\Delta_{n-1} >0$, $\Delta_{n-3} >0$ , 
\end{enumerate}
where $\Delta_{m} $ - denotes the Hurwitz determinant of order $m$ :
\[\Delta _{m} =\left|\begin{array}{cccc} {\begin{array}{c} {a_{1} } \\ {a_{0} } \\ {0} \\ {\begin{array}{c} {0} \\ {\cdot } \end{array}} \end{array}} & {\begin{array}{c} {a_{3} } \\ {a_{2} } \\ {a_{1} } \\ {\begin{array}{c} {a_{0} } \\ {\cdot } \end{array}} \end{array}} & {\begin{array}{c} {a_{5} } \\ {a_{4} } \\ {a_{3} } \\ {\begin{array}{c} {a_{2} } \\ {\cdot } \end{array}} \end{array}} & {\begin{array}{c} {\begin{array}{cc} {\cdot } & {\cdot } \end{array}} \\ {\begin{array}{cc} {\cdot } & {\cdot } \end{array}} \\ {\begin{array}{cc} {\cdot } & {\cdot } \end{array}} \\ {\begin{array}{c} {\begin{array}{cc} {\cdot } & {\cdot } \end{array}} \\ {\begin{array}{cc} {\cdot } & {a_{m} } \end{array}} \end{array}} \end{array}} \end{array}\right| \] 
Obviously, when the Routh-Hurwitz criterion is satisfied, the fixed points are stable  and the position of their equilibrium  is classified as a stable node. We  carry out  numerically analyze of equations (\ref{eq19}) by choosing the values of the parameters  $\textrm{Pm}=1$, $ \textrm{Pr}=9$, $\textrm{H}=5$, $\textrm{T}=1$, $\textrm{Ta}=2$ and $\gamma =1$.  In the case of the Keplerian rotation profile ($\textrm{Ro}=-3/4$)   the critical Rayleigh number: $\textrm{R}_{1cr} \approx 1.4$ is obtained.
If the Rayleigh parameter 
\[\textrm{R}_{1cr}=1+\frac{\textrm{Pm}}{\textrm{Pr}}\textrm{H}+\textrm{T}\sqrt{\textrm{Ta}}\cdot\frac{1+\textrm{Ro}\left(1+\frac{\textrm{Pm}^2}{\textrm{Pr}}\textrm{H}\right)}{1+\frac{\textrm{Pm}}{\textrm{Pr}}\textrm{H}},  \]
then there is one fixed point in the system  $O_{1} (X_{1} ,U_{1} ,Y_{1} ,Z_{1} )$.
Where the critical value of the Rayleigh number   $R_{1cr} $   for stationary convection is:
\[ \textrm{Ra}_{cr}=\frac{a^6}{k^2}+a^2\textrm{Q}+\frac{\pi^2\textrm{Ta}}{k^2}\cdot \frac{a^4+ \textrm{Ro}(a^4+\pi^2\textrm{Q}\textrm{Pm})}{a^4+\pi^2\textrm{Q}}, \]
 that coincides with the expression for  $r$. 

Without taking into account the thermal processes $\textrm{Ra}=0$, the threshold value of the hydrodynamic Rossby number $\textrm{Ro}$ has the form  \cite{10s}-\cite{11s}:  
\[\textrm{Ro}_{\textrm{cr}}  =-\frac{a^2(a^4+\pi^2 \textrm{Ha}^2)^2+\pi^2a^4 \textrm{Ta}}{\pi^2\textrm{Ta}(a^4+\pi^2 \textrm{Ha}^2 \textrm{Pm})}.\]
We calculate the eigenvalues $\lambda _{i} $ as a function of the changes in the Rayleigh parameter $\textrm{R}$  for the second (third) equilibrium state $O_{2,3} $. Here negative stable eigenvalues  $\textrm{Re}\lambda <0$   correspond to stable eigendirections, and positive ones  $\textrm{Re}\lambda >0$   correspond  to unstable directions. The stationary state of convection  $(\lambda=0)$  corresponds to the critical value of the parameter  $\textrm{R}_{2cr} $,  which  turns out to be equal to the first critical value:  $\textrm{R}_{2cr}= \textrm{R}_{1cr} $.

\section{ Conclusion }
                                                                     
Using analytical and numerical methods, we carried out qualitative analysis of a nonlinear system of dynamical equations describing magnetoconvection in non-uniformly rotating electroconductive fluids.
We show the existence of a complex chaotic structure - strange attractor (see Fig. 3). 
It is found that for certain modes of convection with the non-uniformly rotating fluids occur chaotic changes (inversions) of the perturbed magnetic field.
The theory developed in this paper can be used as a scenario for the appearance of turbulence (see Fig. 4) in hot accretion disks.

\end{document}